# An Approximate Feasibility Assessment of Electric Vehicles Adoption in Nigeria: Forecast 2030

Qasim M. Ajao [a,*], and Lanre Sadeeq [a]

[a] *Department of Electrical and Computer Engineering, Georgia Southern University, Statesboro, Georgia, USA*

ABSTRACT

Efforts toward building a sustainable future have underscored the importance of collective responsibility among state and non-state actors, corporations, and individuals to achieve climate goals. International initiatives, including the Sustainable Development Goals and the Paris Agreement, emphasize the need for immediate action from all stakeholders. This paper presents a feasibility assessment focused on the opportunities within Nigeria's Electric Vehicle Value Chain, aiming to enhance public understanding of the country's renewable energy sector. As petroleum currently fulfills over 95% of global transportation needs, energy companies must diversify their portfolios and integrate various renewable energy sources to transition toward a sustainable future. The shifting investor sentiment away from traditional fossil fuel industries further highlights the imperative of incorporating renewables. To facilitate significant progress in the renewable energy sector, it is vital to establish platforms that support the growth and diversification of industry players, with knowledge sharing playing a pivotal role. This feasibility assessment serves as an initial reference for individuals and businesses seeking technically and economically viable opportunities within the sector.

## 1. Introduction:

The ongoing focus on Sustainable Development and the Energy Transition Act is increasingly directing attention toward the transformation of the global energy sector from fossil-based to zero-carbon by the second half of this century. The United Nations, with its commitment to ending poverty, has presented a comprehensive roadmap to safeguard the planet and ensure prosperity for all by 2030 [1, 2]. In response, the Oil and Gas industry is adopting operational models aimed at reducing carbon emissions, while investors are increasingly allocating funds towards opportunities that prioritize high sustainability and societal impact, as guided by the Environmental, Social, and Corporate Governance (ESG) framework. Renewable energy sources play a vital role in the pursuit of universal access to affordable, sustainable, reliable, and modern energy. Among the various applications of renewables—electricity, heat, and transport—renewable electricity has experienced the most significant growth, primarily driven by the rapid expansion of wind and solar technologies. The United Nations has set an ambitious plan in motion, advocating for an integrated approach to achieve swift progress and tangible results [3, 4, 5].

This involves expediting the adoption of proven innovative solutions and establishing partnerships. The UN Climate Action targets for the next decade encompass several key objectives [1]:

- Carbon emissions: Aim for absolute and per capita reductions of 25% by 2025 and 45% by 2030.
- Electricity consumption: Strive for per capita reductions of 20% by 2025 and 35% by 2030.
- Renewable energy: Target a share of 40% by 2025 and 80% by 2030 in terms of consumed electricity.
- Commercial air travel: Work towards per capita emissions reductions of 10% by 2025 and 15% by 2030.
- Climate neutrality: Offset 100% of unavoidable carbon emissions annually from 2019 through certified carbon credits.
- Operational efficiencies: Demonstrate long-term economic benefits through plan implementation.
- Sustainable Development co-benefits: Increase climate-smart infrastructure and sustainable development benefits for local communities resulting from the plan's implementation.

---

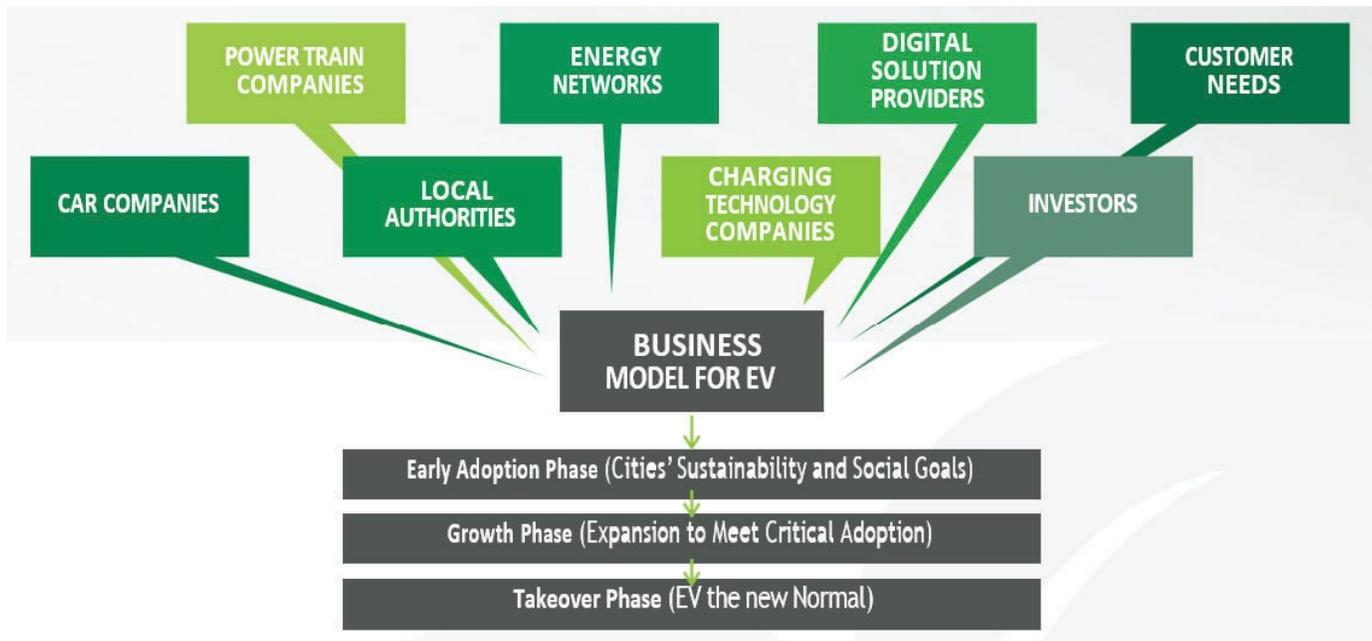

**Fig. 1.** The Architectural System of EV Business

During the first quarter of 2020, global renewable energy usage across all sectors witnessed a 1.5% increase compared to the same period in 2019, indicating that renewable electricity remained largely unaffected while the demand for other energy sources declined [1, 5].

The fluctuating global oil prices, coupled with the global and local shift towards cleaner energy sources, have sparked a change in the preferences of financiers. This transformation has accelerated the need for Nigeria's gas and oil industries to explore diversification strategies toward renewable energy. As the world embraces renewable energy alternatives like solar, wind, geothermal, and tidal power, companies in the fossil fuel sector must swiftly adapt to the evolving market dynamics to avoid disruption. One disruptive element is the increasing adoption of electric vehicles (EVs), which will significantly reduce the demand for traditional fossil fuels. Nigeria, blessed with abundant solar energy resources, is witnessing growing interest from financiers to invest in renewable energy projects across Africa. Add The Nigerian government is also actively promoting solar power initiatives for rural and institutional electrification. This presents promising opportunities for private entities to contribute to Nigeria's environmental sustainability goals. This paper provides an assessment of the Electric Vehicle (EV) value chain, including its technologies, opportunities, and potential obstacles [6].

## 2. Developing the EV Business

The development of the EV business/value chain pertains to the advancement and implementation of technologies aimed at supporting the manufacturing of EV components and the establishment of EV charging infrastructure [5]. The key components within this value chain encompass:

- Manufacturing EV Powertrain and other Sub-Systems.
- Assembly of EV Cars, Distributorship, and Sales.
- Electricity Generation, Transmission, and Distribution Infrastructure.
- Manufacturing of AEVSE and Other EV Charging System Components.
- Charging Infrastructure for EVs (Private and Public).
- E-Mobility Services

The progress of the EV charging business has been sluggish, primarily due to uncertainty surrounding policy direction and timing. This uncertainty discourages investment in potentially stranded assets. To navigate these challenges, investors need to forge partnerships with other stakeholders to collaboratively define the development and growth of the EV industry [7].

Figure 1 depicts the illustrative architectural system showcasing the interconnectedness of various components within the EV business ecosystem.

### 2.1. The EV Charging Business

The progress of the EV charging business has been hindered by uncertainties surrounding policy direction and timing, resulting in a reluctance among investors to invest in potentially stranded assets. To overcome these challenges and drive the development of EVs, investors need to collaborate with other stakeholders to establish a clear roadmap and framework for the industry. Of utmost importance is the need for investors to focus on building infrastructure based on existing demand. Understanding the origins of demand and how consumers will utilize EVs is crucial for effectively sizing, scaling, and shaping the appropriate infrastructure. In addition to the "Home Charging Model," two other models have been defined [6, 7]:

*Mode 1:* The Destination User This model targets areas such as airports, car parks, business parks, and major office spaces where users tend to leave their cars for extended periods of time.

*Model 2:* The Hub User This model focuses on fleets of cars, taxis, buses, emergency vehicles, and delivery trucks. It requires the development of charging hubs strategically located in and around cities. By understanding and catering to these distinct user models, investors can make informed decisions and effectively contribute to the growth of the EV charging infrastructure.

### 2.2. Global Market Outlook

The electrification of transportation is transforming the mobility sector, supported by significant industry changes and trends [3]. Notable observations include:

- Car manufacturers are fully embracing EVs, with an expected introduction of at least 21 new EV brands in 2021 and 25 in 2022.
- Nissan aims to achieve 1 million EV and hybrid sales by FY 2023.
- Renault anticipates EVs to represent 10% of its total sales by 2023, with the Renault Zoe being a popular EV in Europe.
- Daimler plans to include 10 pure electric and 40 hybrid models in its portfolio.
- Volkswagen targets the electrification of its entire vehicle lineup by 2030 and aims for complete CO2 neutrality by 2050.
- Utilities, power companies, and other energy firms have significantly increased investments in EV charging

infrastructure, totaling around $1.7 billion. Moreover, over $100 billion has been allocated for battery and EV cars since 2018.

Political and government support for EVs is significantly growing. The US government has expressed strong support for EV adoption, including plans to establish 500,000 new public charging outlets and reinstate EV tax credits while the UK government has accelerated the ban on fossil fuel vehicles, advancing it to 2030 [8]. Private commercial companies are actively transforming their fleets, including DHL that has committed to achieving 70% clean operations for last-mile pickups and deliveries by 2025, and DB Schenker that also aims to make it activities in Europe emission-free by 2030. As the price gap between internal combustion engine (ICE) vehicles and EVs narrows, estimated to take 2-3 years, these trends act as powerful signals that EVs are here to stay. This exerts pressure on competitors, stakeholders, and investors to act swiftly or risk lagging in this rapidly evolving landscape. The EV market outlook currently presents two prominent scenarios:

1. The State Policies Scenario: This scenario assesses the impact of existing policy frameworks and announced policy intentions on the EV market by 2030. It projects the following by 2030:
   - The global EV stock (excluding two/three-wheelers) will achieve a scale of 140 million.
   - EVs will constitute 7% of the global vehicle fleet.
2. EV30@30 Scenario: This scenario is a part of the Clean Energy Ministerial Campaign, aimed at supporting governments in mitigating greenhouse gas emissions by promoting the sales share of EVs. This campaign sets the following targets for 2030:
   - Global EV sales will capture 30% of global car sales.
   - The annual global sales volume of EVs will reach 43 million.

The realization of these projections relies on several factors, such as the passage of time, the accumulation of new data, and the effective implementation of novel policies. Only through the interplay of these factors will we gain insight into the unfolding of these scenarios. The global electric vehicle fleet has experienced a remarkable expansion over the past decade, driven by supportive policies and advancements in technology. In 2019 alone, 2.1 million electric vehicles were sold, accounting for 2.6% of global car sales. The total number of electric vehicles worldwide reached 7.2 million, representing 1% of the global car stock. Based on the State Policies Scenario, it is projected that the global electric vehicle fleet will reach 140 million by 2030, constituting 7% of the global car stock [3, 6, 9].

Currently, nine countries have surpassed the milestone of having over 100,000 electric vehicles on their roads. While light passenger vehicles remain the most popular type of electric vehicle, there is also a growing trend of adoption among two- and three-wheelers, as well as light public and commercial vehicles. Europe has emerged as a frontrunner in the electric vehicle (EV) market, experiencing consistent growth in EV sales and market share. This achievement has solidified Europe's position as a leader in terms of EV market penetration. The region has shown a strong dedication to transitioning towards sustainable transportation, supported by favorable policies, robust infrastructure development, and increasing interest from consumers. European countries have taken proactive measures by setting ambitious targets and offering incentives such as financial subsidies and tax benefits, and investments in charging infrastructure.

These actions have played a vital role in stimulating consumer demand and driving the widespread adoption of EVs. Moreover, European automakers have made remarkable advancements in producing high-quality electric vehicles, offering a wide range of models that cater to diverse consumer preferences and needs.

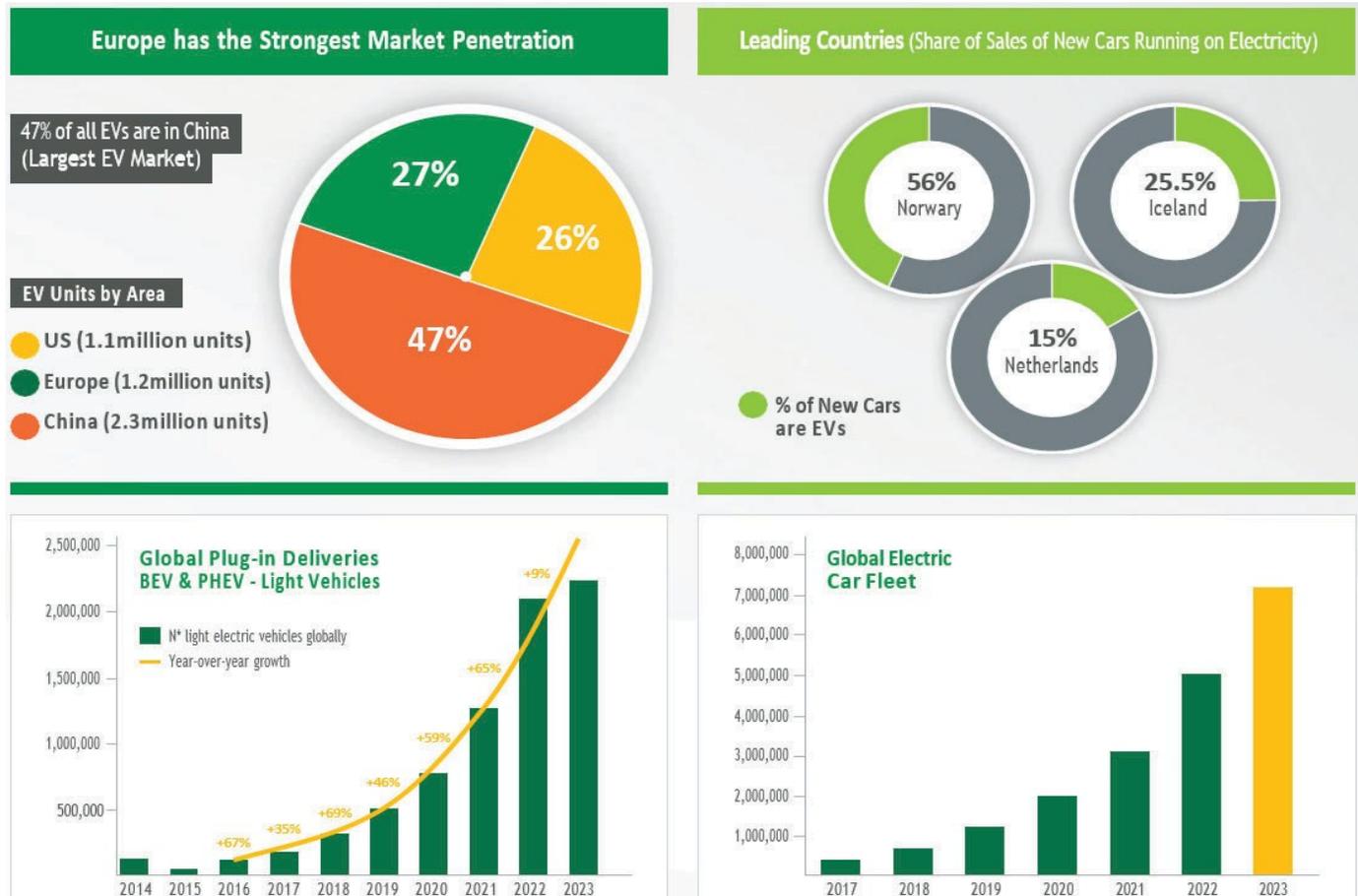

**Fig. 2.** Analysis of Global EV Deliveries and Market Size

The availability of affordable and technologically advanced EV options has further fueled the popularity of electric vehicles across the region. Additionally, collaborations between governments, industry stakeholders, and energy companies have played a significant role in establishing a comprehensive charging network, ensuring convenient and accessible EV charging throughout Europe [10]. These collective efforts have contributed to Europe's success in achieving a remarkable EV market penetration and positioning itself at the forefront of the global EV market.

*2.3. EV Growth and Trends*

The electrification of transportation extends to various vehicle types beyond passenger cars, encompassing two/three-wheelers as well. These two/three-wheelers, including e-scooters, e-bikes, electric mopeds, and electric tricycles, have experienced substantial growth in electric mobility options. They are now available in over 600 cities across 50 countries globally, with notable popularity observed in densely populated regions such as China and India [11]. The integration of electric power in these vehicles has been facilitated by the adoption of the battery swap method for recharging, offering a convenient and efficient solution [5]. In addition to two/three-wheelers, the electrification trend is also evident in the domain of light commercial vehicles. Companies and public authorities are increasingly incorporating electric models into their fleets, contributing to the electrification of transportation in this segment. Furthermore, electric buses are gaining traction as countries worldwide strive to electrify their public transportation networks. An exemplary case is Chile, which has set a target of achieving full electrification by 2040 [8]. These developments underscore the growing acceptance and integration of electric vehicles in a diverse range of transport modes, extending beyond the traditional focus on passenger cars.

## 3. EVs in Africa and Nigeria

A survey conducted by auto-trader reveals that electric cars have piqued the interest of Africans, but their high cost remains a significant barrier for most car owners. The survey sheds light on the factors that drive or impede the adoption of electric vehicles (EVs) in Africa. South Africa took the lead in EV adoption by introducing the Nissan Leaf in 2014, and the country currently boasts an estimated 1000 EVs [9]. Nairobi, Kenya, Uganda, Rwanda, and Nigeria also have EVs in circulation. However, EVs only represent a mere 0.001 percent of car sales in Africa, with the predominant adoption strategy being their utilization of ride-hailing services. Regarding charging infrastructure, South Africa has made remarkable strides and now possesses the most advanced charging infrastructure in Africa, with investments exceeding 2 million USD in the electric power way project [7]. A paradigm shift in mindset is imperative to foster widespread EV adoption across Africa [12]. The Middle East and Africa region is projected to experience a Compound Annual Growth Rate (CAGR) of approximately 6.80% from 2020 to 2025, with Dubai aiming to achieve a 30% share of EVs in road transport by 2030. In Nigeria, Hyundai and Stallion Group have taken a significant leap forward in EV deployment and adoption by introducing the first locally assembled EV electric car, equipped with a 64-kWh battery pack that enables a range of 300 miles (482 km) on a single charge.

The introduction of EVs in Nigeria presents both challenges and opportunities. Given the prevailing power situation in the country, several questions arise concerning the power source, distribution of generated power, and charging infrastructure. EV owners must contemplate charging options at home or public stations while addressing the ownership and operation of public charging infrastructure [13, 14]. It is crucial to find solutions to these inquiries to surmount the unique challenges posed by the Nigerian business environment and foster the widespread adoption of EVs in the country [15]. The EV potential in Africa currently presents two prominent scenarios:

1. *The Drivers:*
   - The anticipated fuel savings of 40-70% motivate the adoption of electric vehicles.
   - The rapid growth of urbanization creates opportunities for electric mobility.
   - Micro-mobility and gig economies offer new possibilities for electric vehicle usage.
   - Lower lifetime running costs compared to internal combustion engine (ICE) vehicles attract potential EV owners.
   - Overcoming fossil fuel scarcity, especially in Sub-Saharan Africa (SSA), drives the transition to electric vehicles.
   - Environmental concerns and a desire for greener mobility contribute to the EV demand.
   - The reduction in noise pollution is a positive aspect of electric vehicles.

2. *The Resistors:*
   - Impending global regulations may impact local markets and stem challenges to EV adoption.
   - Higher upfront costs, including high import tariffs and a lack of subsidies, hinder EV affordability.
   - The current lack of charging infrastructure limits the widespread use of EVs.
   - EV range limitations can cause range anxiety and discourage potential buyers.
   - High electricity prices affect the cost-effectiveness of electric vehicle ownership.
   - Instability in grid electricity, particularly due to load shedding, poses obstacles to EV charging.
   - Long charging times may inconvenience EV users and affect their overall experience.
   - The lack of enabling policies, such as tax incentives and subsidies, hinders the growth of the EV market.
   - Political will to support EV infrastructure development creates barriers to progress.
   - Existing ICE vehicles still have usable life and are perceived as reliable, leading to reluctance in transitioning to EVs.

The electric mobility market in Africa is anticipated to experience a substantial surge in growth, especially in the segment of two and three-wheelers. This positive outlook is attributed to several factors. Firstly, the increasing urbanization in various African countries is driving the demand for efficient and sustainable transportation solutions [9]. Electric two and three-wheelers offer a viable alternative to conventional vehicles, particularly for short-distance commuting within crowded urban areas. Furthermore, the initiatives supported by the United Nations in countries like Ethiopia, Morocco, Kenya, Rwanda, and Uganda are contributing to the expansion of the electric mobility market. The United Nations' involvement provides credibility, resources, and guidance to local governments and organizations, encouraging them to invest in electric mobility infrastructure, develop supportive policies, and incentivize the use of electric vehicles [10, 15].

Leading automotive brands have recognized the potential of the African market and have introduced their electric vehicle models. Notable examples include the BMW Mini-Cooper SE, Jaguar I-Pace, Nissan Leaf, BMW i3, Volkswagen E-Golf, Hyundai Kona, and Hyundai Ioniq. This diverse range of offerings caters to the varying preferences and needs of consumers, further fueling the growth of the electric vehicle market in Africa. To bolster the development of the EV industry in Africa, several EV assembly plants have been established between 2018 and 2020. These plants, located in Kampala (Uganda), Kigali (Rwanda), Lagos (Nigeria), and Addis Ababa (Ethiopia), demonstrate a growing commitment to local production and assembly of electric vehicles within the continent. This not only supports the expansion of the EV market but also contributes to job creation and economic growth in these regions [11, 16].

*3.1. Nigeria's Capability and Sustainability*

The current state of electricity accessibility in Nigeria, with a 60% access rate and a national grid dependent on load shedding, presents a significant challenge for the development of electric vehicles in the country.

In addition to cost considerations, the inadequate energy infrastructure poses a major barrier to the widespread adoption of EVs in Nigeria. Initially, EV ownership in Nigeria may be limited to affluent individuals who can generate their own electricity through sources such as PHCN, diesel/petrol generators, and potentially solar power for Level 1 or Level 2 home charging [10]. Alternatively, they may need to pay a premium for charging at privately or government-owned public charging stations, depending on the capacity of the existing power supply. To establish EVs as a viable transportation solution on a large scale in Nigeria, several factors must be addressed. Firstly, there is a pressing need to upgrade and expand power generation, transmission, and distribution capacity to meet the growing demand for electricity from EVs. Moreover, the affordability of EVs must be improved in comparison to conventional fossil fuel-powered vehicles. The cost of power per distance traveled should also be more economically advantageous than the cost of fuel per distance covered. Additionally, the integration of gas-powered and solar EV charging stations into the energy mix is essential to ensure a sustainable and efficient charging infrastructure [15].

*3.1.1. Key Rist Indicators and Mitigation Measures*

Limited experience in the sector and lack of local technical expertise poses significant risks to the development of the electric vehicle industry. There is a lack of knowledge and expertise required to effectively develop, produce, replicate, and control the technological principles in the product and service, particularly in the areas of Electric Vehicle Supply Equipment Supplier (EVSE-S) and Charge Point Operator (CPO) operations, as well as E-mobility Service Provision. To mitigate these risks, it is essential to seek working partnerships and technical alliances with renowned international players in the electric vehicle industry. Collaborating with established entities will help augment local skill sets, acquire new competitive skills, and enable technology and knowledge transfer, which will have a lasting impact on the brand's positioning in the market [17].

Another risk factor is the slow development of the EV charging business, driven by uncertainty surrounding policy direction, timing, and the inherent technical limitations of range (One-Time Travel Distance at Full Charge). This limitation can potentially cause range anxiety, especially for local long-distance travelers [5, 15]. To address this risk, it is important to actively drive policy changes and support within the electric vehicle sector. Building a robust and stable policy framework, along with long-term national objectives and targets, supported by sound market forecasts, will instill investor confidence. Policy approaches can include incentives for zero- and low-emissions vehicles, economic instruments to bridge the cost gap between electric and conventional vehicles, and support for the phased deployment of charging infrastructure. Furthermore, increasing the number of charging stations, in the long run, can help mitigate the limited range problem. Technological advancements have also introduced the battery swap method of recharging, which reduces charging time and provides an alternative solution. By implementing these mitigation measures, such as seeking partnerships, driving policy changes, and expanding charging infrastructure, the risks associated with limited experience, technical expertise, and range limitations can be effectively addressed and pave the way for the sustainable growth of the electric vehicle industry [18].

*3.1.2. Feedstock Resources*

Inadequate local electricity supply and infrastructure pose a significant risk to sustaining the Electric Vehicle (EV) business/industry in Nigeria. The country's low electricity access rates and reliance on load shedding within the national grid complicate the emergence of EVs [19]. To mitigate this risk, it is crucial to build infrastructure tailored to meet the existing demand. This requires a thorough understanding of current and potential demand, enabling the strategic sizing, scaling, and shaping of infrastructure. A phased approach can be adopted to introduce home, office, and public charging models, accommodating the diverse needs of EV users [16].

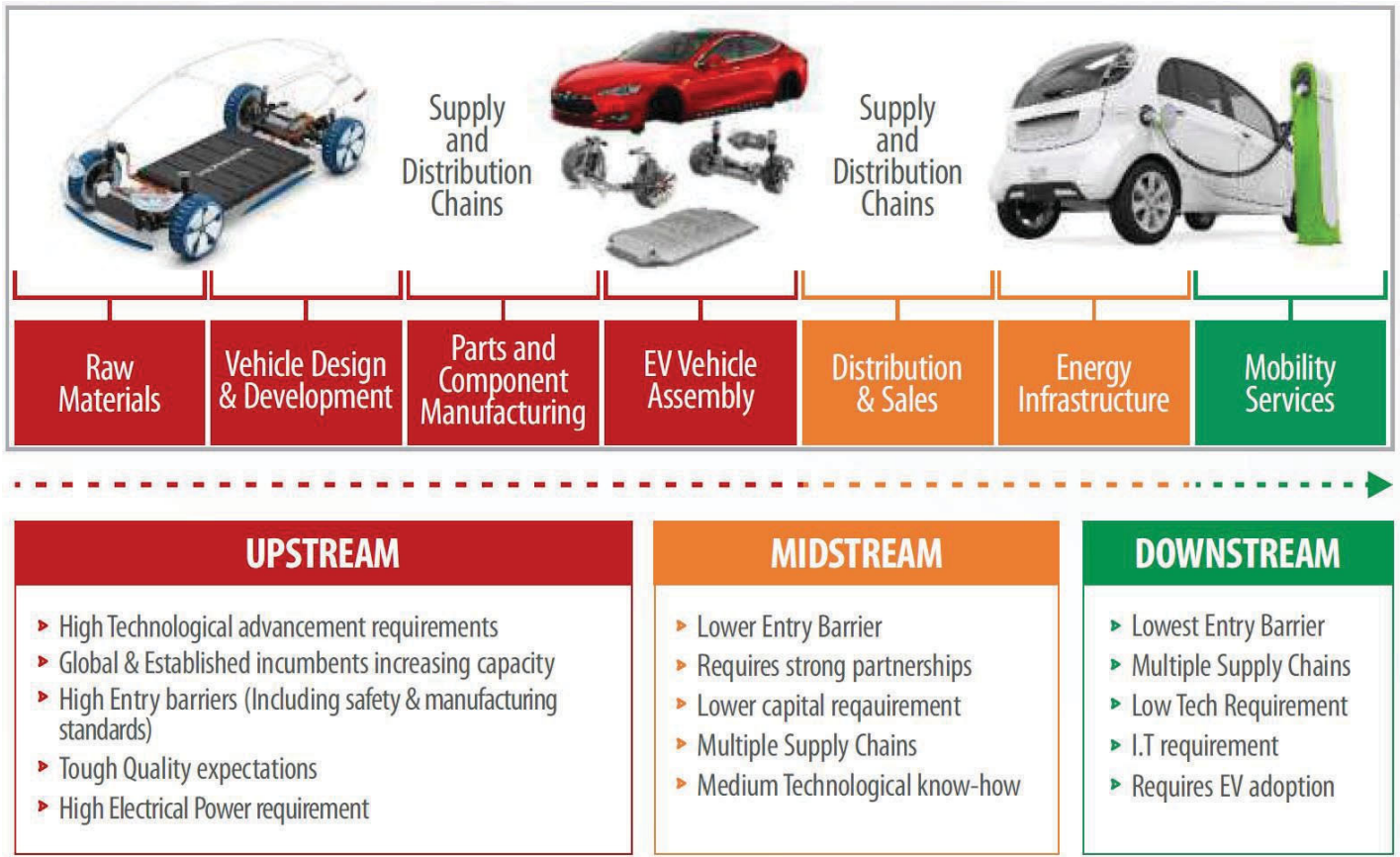

**Fig. 3.** Market Entry Strategy and Approach

Furthermore, initial adopters can address the electricity supply challenge by generating their power through sources like The Nigerian Electricity Regulatory Commission (NERC), diesel/petrol generators, or solar energy for Level 1 or Level 2 charging at home. Alternatively, they can opt for charging at private or government-owned public stations, provided the existing power supply can support it. Upgrading and expanding power generation, transmission, and distribution capacity is essential to facilitate widespread EV adoption as a reliable means of transportation in Nigeria. Additionally, the affordability of EVs plays a significant role in their adoption. Mitigation measures should focus on making EV pricing comparable to fossil fuel-powered vehicles. Moreover, the cost of power per distance traveled must be more economical than the cost of fuel per distance traveled. This can incentivize consumers to choose EVs over conventional vehicles [5, 7, 13]. Diversifying the energy source mix by incorporating gas-powered and solar EV charging stations can also enhance the charging infrastructure and provide more options for EV owners. Another risk to consider is the limited availability of public charging stations, which hampers widespread EV adoption due to insufficient charging networks. To address this, building infrastructure based on existing demand becomes critical. By strategically sizing, scaling, and shaping the infrastructure, it becomes possible to meet the charge needs of EV users. Encouraging office policies that promote EV adoption as official vehicles, aligned with ESG sustainability adoption by public companies, can further drive the establishment of more charging stations. Technological advancements such as battery swap methods, particularly suitable for two- and three-wheelers, can reduce charging time and facilitate easier adoption [20].

Supply chain risks pose additional challenges in the EV industry. To mitigate these risks, it is essential to leverage technical partnerships with component manufacturers, strengthening the supply chain and ensuring reliable access to necessary components. Building strategic relationships and thoroughly assessing supply chain partnerships allow for supply optionality and the availability of alternative backup suppliers. Implementing quality control measures, such as in-line and pre-shipment inspections, ensures the components' reliability and adherence to standards. Maintaining module/component delivery timelines through a risk-based logistics strategy further mitigates potential disruptions in the supply chain. The risks associated with inadequate electricity supply, limited charging infrastructure, and supply chain complexities can be mitigated through strategic infrastructure development, policy support, technological advancements, and effective supply chain management. By addressing these risks and implementing appropriate mitigation measures, Nigeria can pave the way for a sustainable and thriving electric vehicle (EV) industry, positioning itself as a leader in the adoption of clean and efficient transportation solutions [21].

*3.1.3. Output and END Use*

The slow adoption of electric vehicles (EVs) due to consumer perceptions is a significant risk indicator that needs to be addressed in the e-mobility industry. Concerns about infrastructure to support EV adoption and the limited driving distance with a single charge are key factors contributing to this slow adoption [19, 22]. Moreover, the higher pricing of EVs compared to conventional vehicles (CVs) acts as a deterrent for potential buyers. The influence of social factors, particularly consumer understanding of the benefits and attributes of EVs, cannot be underestimated in shaping users' preferences between EVs and CVs. To foster greater adoption, it is crucial to focus on sensitizing and educating potential EV users about the high quality, environmental advantages, and long-term financial savings associated with EV ownership, such as reduced maintenance costs. In making well-informed investment decisions within the e-mobility value chain, the assessment of optimal profitability becomes paramount. Comprehensive economic and financial models should be employed to evaluate the profitability of different streams within the value chain. This evaluation provides stakeholders with the necessary information to make final investment decisions that align with their objectives. By considering these risk indicators and implementing effective mitigation measures, including consumer education initiatives and thorough profitability evaluations, the adoption of EVs can be actively encouraged and supported.

These steps are critical for a sustainable and successful transition to widespread EV usage, contributing to a greener and more efficient transportation system [11, 15, 23].

*3.2. Economics and Financing*

Overcoming the challenges of consumer perceptions, limited infrastructure, and higher pricing will create an environment conducive to widespread EV adoption. This will not only drive economic growth and job creation in the EV sector but also contribute to reducing carbon emissions, enhancing energy security, and promoting sustainable development in Nigeria [24]. The relatively higher price of electric vehicles (EVs) compared to conventional vehicles (CVs) presents a significant local and regional barrier, impacting market penetration, demand, and profitability. This limitation is influenced by the slow rate of EV adoption in Nigeria and Africa, along with the higher cost of electric charging and EV component replacement [10]. To overcome this challenge, it is essential to drive the implementation of economic policies and incentives that narrow the cost gap between EVs and CVs. Supporting early charging infrastructure deployment and implementing policy measures like parking waivers or reduced toll fees can enhance the value proposition of EVs and foster their adoption [25].

Gaining an in-depth understanding of current and potential demand is crucial to drive government and office policies for adopting EVs as official vehicles, aligning with Environmental, Social, and Governance (ESG) sustainability principles [7]. Raising public awareness about the environmental benefits and long-term financial advantages of EV adoption, such as lower maintenance costs and reduced carbon emissions, is also vital. To determine optimal strategies for achieving market penetration, demand, and profitability across various value chain segments, detailed economic and financial models need to be developed. These models will inform the final investment decision and guide project commercialization efforts. Securing sufficient funding and financial capabilities poses another significant challenge. Identifying local and international intervention funds and grants and positioning oneself to access them is crucial. A thorough assessment of eligibility criteria for identified funds and grants should be conducted, and strategic positioning to meet those criteria is necessary. In cases where prequalification poses time or experience-based barriers, forming partnerships or technical alliances with companies that meet the requirements can be advantageous [26]. Addressing alternative funding barriers and mitigating the perceived high cost of doing business in Nigeria is vital to maximize the project's value creation potential. Developing a business model focused on cost optimization and profitability, particularly in the commercialization of energy or power output, is essential. Conducting a comprehensive project evaluation and commercial optimization analysis is crucial to ensure sustainability and profitability. By proactively addressing these risks and implementing appropriate mitigation measures, Nigeria can overcome barriers to EV adoption and pave the way for a sustainable and thriving EV industry. This will not only contribute to economic growth and job creation but also foster environmental sustainability and energy efficiency [18, 21, 27].

*3.3. Government and Regulatory*

Insufficient policy support and regulatory uncertainty pose significant challenges for major investors and entrepreneurs in the electric mobility industry. Presently, Nigeria lacks tax credits for renewable energy, as the government is in the process of formulating comprehensive policies to incentivize and promote the development of solar power and renewable energy [28]. To foster the expansion of electric mobility in Nigeria, the government must implement a range of measures. These include revitalizing the electricity supply infrastructure, establishing procurement programs to stimulate demand and encourage automakers to increase the availability of EVs in the market, offering incentives for the initial deployment of publicly accessible charging infrastructure, enforcing fuel economy standards, and providing economic incentives for zero and low-emission vehicles [29]. Additionally, it is crucial to bridge the cost gap between electric and conventional vehicles through economic incentives, while supporting the early deployment of charging infrastructure and implementing other policy measures that enhance the value proposition of EVs, such as parking waivers or reduced toll fees. Policy support should be extended to acknowledge the strategic importance of the EV technology value chain [30]. To instill confidence among investors, a robust and stable policy framework with long-term national objectives and targets, underpinned by reliable market forecasts, is indispensable. Efforts should be made to advocate for policy changes and garner support within the electric mobility sector. These endeavors will contribute to creating a conducive investment environment, attracting financial investments, driving technological advancements, and fostering sustainable transportation and environmental objectives. Addressing these risks and implementing appropriate mitigation measures, can assist Nigeria to overcome policy-related challenges and establish an enabling environment for the growth of the electric mobility industry. This will not only attract investments but also stimulate technological innovation, advancing the country's transportation landscape and contributing to sustainable development goals [20, 23].

*3.4. Carbon Credits in Nigeria*

Carbon credits in Nigeria are an integral part of the global framework established by the Kyoto Protocol, with the Clean Development Mechanism (CDM) playing a crucial role in developing countries. Within this framework, Annex B Countries have the ability to provide financial support for emissions reduction projects in Nigeria and earn certified emission reduction (CER) credits in return. To qualify for CDM, projects in Nigeria must showcase long-term climate change benefits and achieve emissions reductions that surpass what would naturally occur. The administration of carbon credits in Nigeria is overseen by the Presidential Implementation Committee for CDM, operating under the Federal Ministry of Environment. Nigerian companies actively involved in reducing greenhouse gas (GHG) emissions are eligible for carbon credits, which they can sell to companies surpassing their emission allowances, thereby raising funds. It is noteworthy that income generated from carbon credit trading in Nigeria enjoys tax exemption. The price of carbon credits in Nigeria, akin to other markets, is influenced by factors such as demand, supply, and risks associated with projects, sovereign entities, and credit [31].

*3.5. The CBN Intervention Funds*

The Central Bank of Nigeria established the CBN intervention funds in January 2016 to provide financial support to strategic subsectors such as agriculture, manufacturing, mining, solid minerals, and others. These funds have a specific focus on supporting both green projects and brown projects (expansion projects) that prioritize local content, foreign exchange earnings, and job creation. It should be noted that trading activities are not eligible for these intervention funds. Within the upstream sector, there are two types of funding available: a Term Loan for acquiring plants and machinery and Working Capital financing. The maximum repayment period for the Term Loan is 10 years, while the Working Capital loan has a one-year tenor with the option to roll over annually. The interest rate for these intervention funds is set at 9% [32].

Additionally, there is a one-year moratorium period during which the borrower is not required to make repayments. To qualify for the CBN intervention funds, borrowers must be registered under the Companies and Allied Matters Act (CAMA). This requirement ensures that the funds are directed toward legitimate businesses that adhere to the necessary regulatory requirements [31].

**4. EVs in Africa and Nigeria**

Electric Vehicles (EVs) are classified into four main categories: Battery Electric Vehicles (BEV), Extended Range Electric Vehicles (EREV), Plug-In Hybrid Electric Vehicles (PHEV), and Hybrid Electric Vehicles (HEV). These vehicles utilize electric motors instead of internal combustion engines for propulsion. EVs primarily rely on battery packs to store electrical energy, which powers the electric motor. Charging the EV battery is achieved by connecting the vehicle to an external electric power source. It is noteworthy that BEVs are recognized as Zero-Emission vehicles by the U.S. EPA due to their lack of direct exhaust or tailpipe emissions, despite potential air pollution during the charging process [5, 33].

The powertrain of an EV encompasses the components responsible for generating and delivering power to the wheels, including the battery, electric motor, and associated systems (refer to Figure 4 and Figure 5). In contrast, gasoline vehicles rely on spark-ignited internal combustion engines, fuel systems, transmissions, and electronic control modules (ECM) to manage fuel mixture, ignition timing, emissions, operations, and diagnostics.

***Hybrid EVs*** combine spark-ignited internal combustion engines with electric generators that charge the traction battery during braking. They also feature electric traction motors to power the vehicle at low speeds or during idle periods. Similar to gasoline vehicles, hybrid EVs include fuel systems, transmissions, and power electronics controllers to manage the flow of electrical energy [34].

***Plug-in hybrid EVs (PHEVs)*** differ from HEVs in that their traction battery packs can be charged through various methods, such as regenerative braking, wall outlets, charging equipment, and the internal combustion engine. PHEVs possess larger battery packs, onboard chargers, and charging ports [35].

***Battery Electric Vehicles (BEVs)*** operate exclusively on electricity stored in an onboard traction battery pack. They depend on external electrical power sources for charging and lack internal combustion engines or fuel systems. BEVs are equipped with scaled-up electric traction motors compared to PHEVs or HEVs [34, 35].

***Hydrogen fuel cell vehicles*** employ hydrogen fuel cells to generate electricity, which powers the electric motor. The fuel cell stack consists of individual membrane electrodes that facilitate an electrochemical reaction between hydrogen and oxygen, resulting in electricity production with water as a byproduct [35].

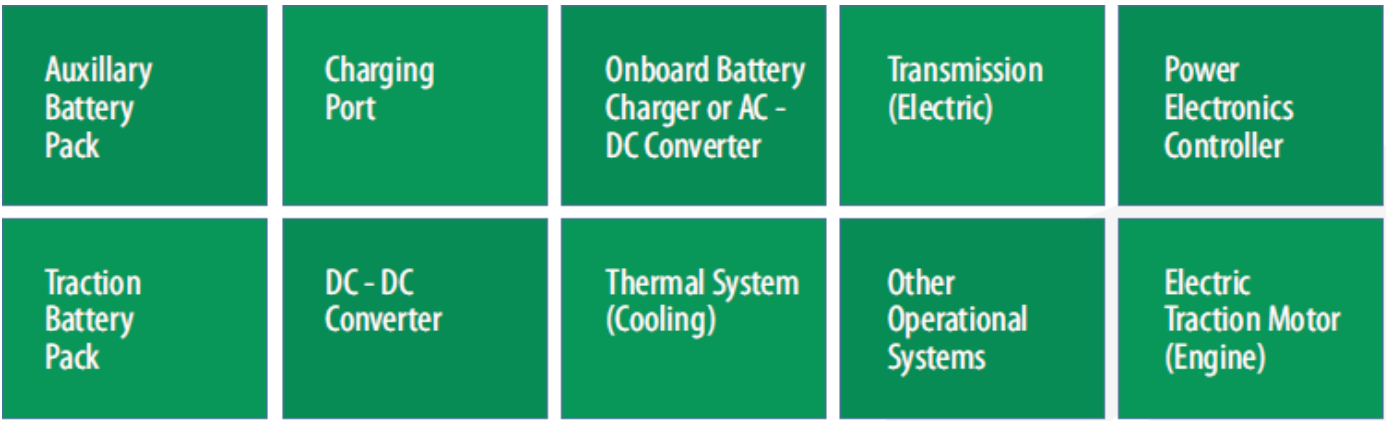

**Fig. 4.** Key Components and Systems of an Electric Vehicle

*4.1. EV Part Manufacturing in Nigeria*

In the manufacturing of Electric Vehicles (EVs), two main systems require specific production processes: the electric motor and controller system, and the battery storage system. The electric motor and controller system play crucial roles in the operation of an EV. The controllers are responsible for managing the voltages and currents that flow from the external electric supply, the battery, the electric motor, and other systems within the vehicle. Electric motors, on the other hand, convert electrical energy into mechanical motion for propulsion. These systems are typically designed by car companies to be manufactured either in-house or by third-party manufacturers. The battery storage system is another essential component of an EV. It consists of several connected battery cells enclosed in specially designed housing, which is often integrated into the chassis of the electric vehicle. The battery cells are usually purchased from battery manufacturers by the EV manufacturer, based on the required dimensions that allow for easy configuration and scalability [36].

These battery cells are then assembled into modules and incorporated into the battery pack, which serves as the primary energy storage for the vehicle. In the context of feasibility in Nigeria, the establishment of local manufacturing capabilities for EV parts, including electric motors, controllers, battery cells, and battery packs, would be crucial [10]. This would require collaborations between car companies, battery manufacturers, and other relevant stakeholders to develop the necessary expertise and infrastructure for local production. By promoting domestic manufacturing, Nigeria can reduce reliance on imported EV components, create job opportunities, and contribute to the growth of the electric mobility industry in the country [14].

*4.2. EV Charging and Energy Infrastructure*

The development of EV energy infrastructure involves implementing technologies to support electric vehicle charging across various applications. This infrastructure includes elements such as electricity generation, transmission, and distribution systems, as well as both private and public charging infrastructure and smart metering solutions [37]. In this development, various stakeholders play important roles. These stakeholders include energy suppliers or generation companies (GENCOs), distribution companies (DISCOs), charge point operators (CPOs), charge location owners, mobility service providers (MSPs), roaming platform providers, and electric vehicle drivers or fleet managers. When it comes to EV charging, electric cars require their batteries to be charged after use. This charging process is facilitated by Electric Vehicle Supply Equipment (EVSE), which conditions and transfers energy from the electrical supply network to the EV's traction battery for charging [38].

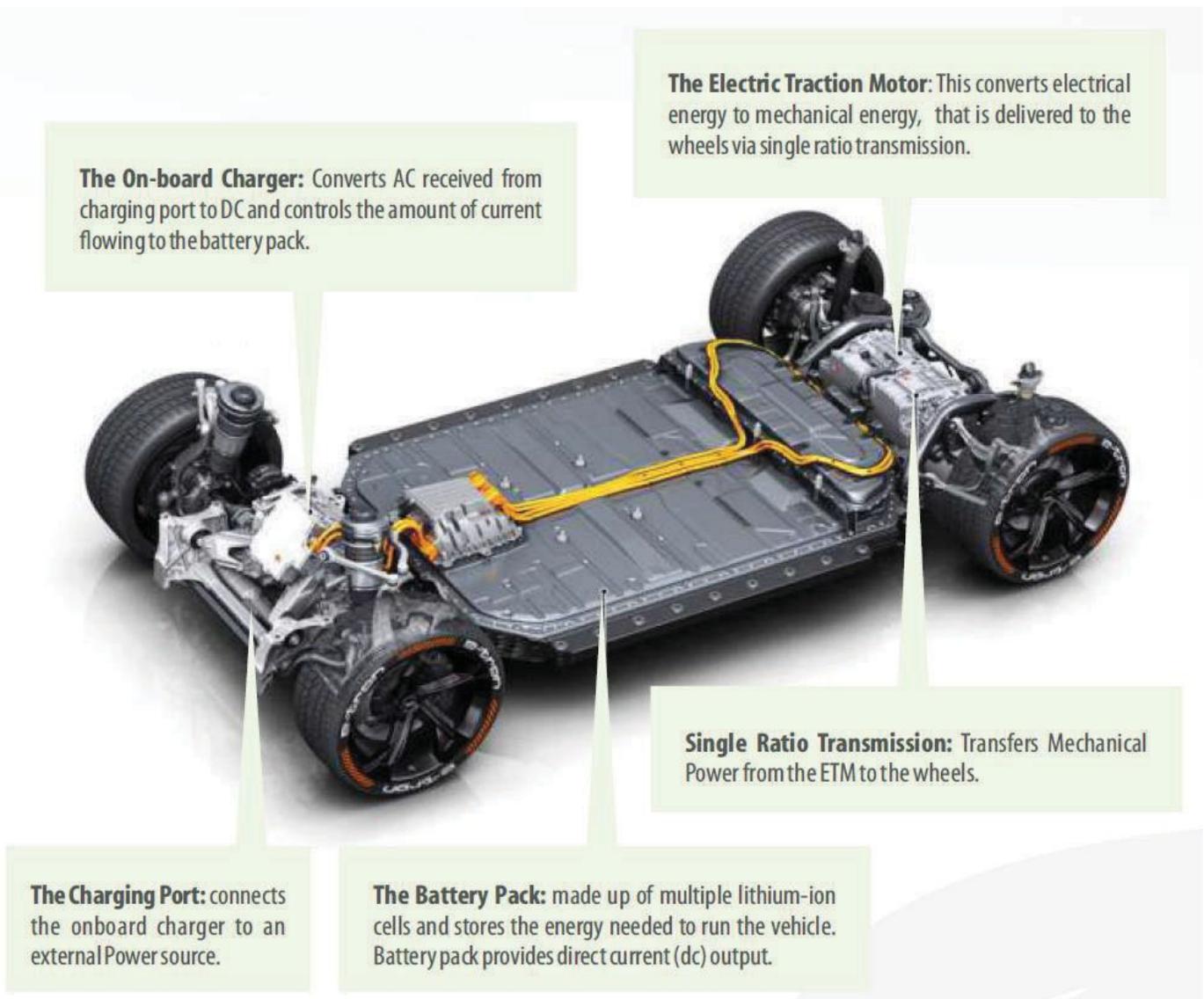

Fig. 5. Power Train Components of an Electric Vehicle

There are generally three methods of charging:
1. Conductive Charging: This method involves connecting the EV battery to an electricity source or charging unit using a cable. Conductive charging is further classified into three levels:
   - Level 1 Charging (Home/Public): Utilizes a 120V power supply.
   - Level 2 Charging (Home/Public): Utilizes a 240V power supply.
   - Level 3 Charging (Public): Requires a 480V power supply and is commonly found in public charging stations.
   - Battery Swapping: This method involves swapping discharged EV batteries with fully charged ones at designated swapping stations.
2. Inductive Charging: Inductive charging enables charging without physical contact between the EV and the charging infrastructure through electromagnetic transmission. Inductive charging can be further classified into two types:
   - Static Inductive Charging: Involves stationary charging pads or plates installed at specific locations.
   - Charging Lanes: Incorporates charging technology embedded in the road, allowing for dynamic charging while driving.

Developing EV energy infrastructure and charging capabilities is crucial for facilitating the widespread adoption of electric vehicles. It requires collaboration among various stakeholders, including energy providers, charging infrastructure operators, and government agencies, to establish a reliable and accessible charging network that supports the growing EV market [39].

*4.3. EV Conductive Charging and Levels*

The conductive charging system utilizes a direct physical connection between the EV connector and the charge inlet. To charge the EV, the cable is connected to either a standard electrical outlet or a dedicated charging station. However, one limitation of this method is the manual requirement for the driver to physically plug in the cable, which is primarily a matter of establishing the connection [7, 8, 33]. Conductive charging encompasses different charging levels, which correspond to the power capacity of the charging outlet. In the Nigerian context, as the popularity of EVs increases and their charging efficiency and speeds improve, it is anticipated that EV owners will prefer to charge their vehicles at home using either Level 1 or Level 2 chargers. Home charging offers advantages in terms of convenience and cost-effectiveness, as charging at home is generally more economical compared to using public charging stations [39].

Level 1 home chargers are directly connected to standard wall sockets and provide a relatively low charging rate, typically measured in range per hour. These chargers are commonly supplied with the EV during purchase and take approximately 20 hours to fully charge an EV within a 200 km range. The cost of charging with Level 1 chargers depends on the electricity tariff applicable in the owner's location. On the other hand, Level 2 home chargers deliver a faster charging rate, around 4-10 times quicker than Level 1 chargers. They provide a charging rate of 12-60 miles per hour. Level 2 home chargers are sold separately from the EV and necessitate specialized installation services by original equipment manufacturers (OEMs) or certified electricians. These chargers typically operate at a rating of 240V and 16A charging current. The cost of Level 2 chargers can vary from $500 to $800, with installation costs ranging from $1,000 to $3,000, including necessary permits [38, 39].

Public charging stations cater to EV drivers who require longer-distance travel beyond the full range capacity of their EVs. These stations are commonly located in public areas such as restaurants, shopping centers, and office parking lots. Similar to Level 2 home chargers as depicted in Figure 7, Level 2 public chargers provide an increased charging rate, ranging from 12-60 miles per hour. The average installation costs for Level 2 public chargers typically range from $1,000 to $4,000. Level 3 public chargers, also known as DC fast chargers, offer a significantly faster charging rate compared to Level 1 and Level 2 chargers [40].

They are approximately 20- 40 times faster than Level 1 chargers and 8-10 times faster than most Level 2 chargers. However, DC fast chargers are more costly and require specialized installation services. These chargers are primarily intended for commercial applications and are not intended for residential use. A typical Level 3 charger has a rating of 50 kW with a voltage of 480V, enabling the charging of a 200 km range EV within 30-45 minutes. The price range for Level 3 chargers falls between $10,000 and $50,000, while installation costs for dual-port chargers vary from $4,000 to $20,000, depending on the availability of existing infrastructure (refer to Figure 7 for the Dual-Port Level 2 Public Charger). It's essential to note that not all EV models are compatible with Level 3 chargers, as they require unique charging connectors and specific powertrain architecture [33 – 40]. Figure 6 as shown below illustrates the AC levels and attributes of the DC level.

*4.4. EV Charging Connectors*

It is essential to consider the charging connectors when discussing EV charging. Much like charging cables for mobile phones, EV charging cables typically have two connectors: one that plugs into the vehicle socket and another that connects to the charge point. However, certain charge points may utilize "tethered" charging connectors, where the cable is permanently attached to the charge point. The specific type of connector required depends on the vehicle and the power rating, or "speed," of the charge point. For slow or fast charging, electric vehicles are equipped with either a Type 1 or Type 2 socket. Additionally, they may feature CHAdeMO or CCS connectors for DC rapid charging. In most cases, slow or fast charge points incorporate a Type 2 socket, occasionally accompanied by an attached cable. Conversely, all DC rapid charging stations utilize tethered cables, predominantly equipped with either a CHAdeMO or CCS connector [41].

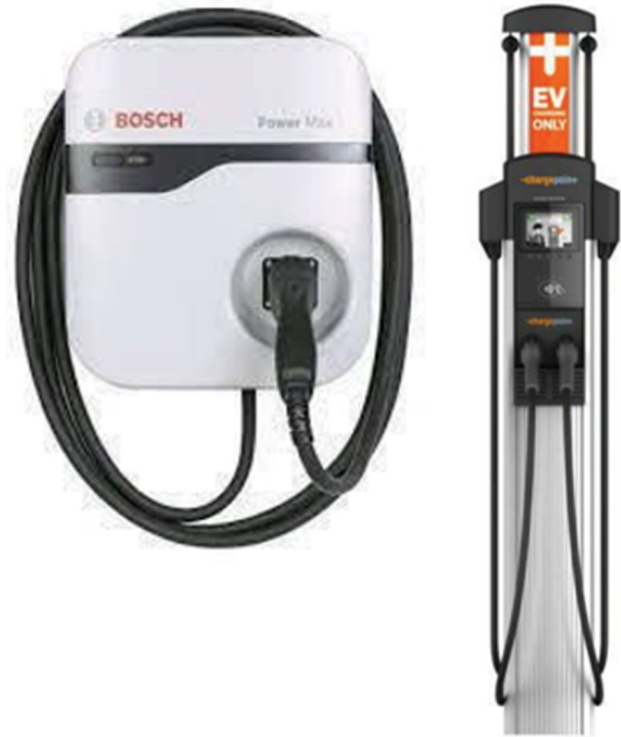

**Fig 6.** The Bosch Power Max Level 2 Home Charger and A Dual Port– Pedestal Mounted Level 2 Public Charger

*4.5. Inductive Charging and Battery Swapping*

Inductive charging employs electromagnetic fields to transfer energy between two objects by transmitting electricity through an air gap from a charger's magnetic coil to a car's fitted magnetic coil [39]. When the vehicle is properly positioned to align both coils, the charging process begins, utilizing the proximity-based induction coils as an electrical transformer [42].

Advanced inductive chargers such as Qualcomm's Halo, as well as those developed by BMW and Tesla, offer a high-quality charging experience. This technology boasts an impressive efficiency level, with only around 10% power loss during charging. Inductive charging pads can be purchased and installed on most new EVs, typically requiring professional installation and costing between $1,500 to $3,000. On the other hand, battery swapping involves the exchange of a depleted battery pack with a fully charged one, eliminating the waiting time for battery recharging. This swap typically occurs at dedicated battery swapping stations (BSS). While battery swapping has encountered challenges in the past, including limited adoption by car manufacturers and gas stations, recent developments have emerged. For instance, the Chinese company NIO has established 125 battery swapping stations for its electric vehicles, offering free battery swaps as an incentive for customers. However, building and maintaining battery swapping stations can be expensive, and the cost of battery replacements often falls on the manufacturers. It is important to note that as EV range and charging times continue to improve, battery-swapping technology is expected to be gradually phased out in favor of more efficient charging methods [43].

*4.6. Smart Charging and the Benefits*

As depicted in Figure 8, smart EV charging provides a multitude of features and advantages for municipalities and utility operators. A key feature is the seamless availability of charging stations for electric vehicles, facilitated by the integration of intelligent technology. EV owners can access real-time information about the location and accessibility of charging stations, enabling them to plan their charging sessions with greater efficiency [44]. This not only enhances convenience for EV owners but also optimizes the utilization of charging infrastructure, reducing congestion at charging stations. Another significant benefit of smart EV charging is the capability to expedite vehicle charging through the implementation of advanced Battery Management Systems (BMS). These systems employ monitoring and regulation of charging parameters to ensure efficient and safe charging. By leveraging intelligent charging algorithms, BMS can deliver the maximum charging power that the vehicle and battery can safely handle, thereby reducing overall charging time for EVs [20, 24, 39, 40].

Safety is of paramount importance in EV charging, and smart charging technologies provide enhanced safety features. Real-time monitoring and diagnostics enable smart charging systems to detect anomalies or faults in the charging process, ensuring that charging sessions are conducted safely and mitigating potential hazards. This instills confidence in EV owners and promotes the widespread adoption of electric vehicles. A significant advantage of smart EV charging is the potential for cost savings [42].

Utility operators and municipalities can offer network incentives, discounts, and benefits through smart charging platforms. These incentives may include dynamic pricing models that encourage off-peak charging when electricity rates are lower, as well as rewards programs for participating in demand response initiatives [43]. By optimizing charging patterns and leveraging the flexibility of smart grids, EV owners can benefit from reduced charging costs, leading to overall financial savings.

Smart EV charging also contributes to grid stability and effective energy management. Through remote control and management of charging sessions, utility operators can ensure that charging activities are aligned with grid availability and stability [44]. This becomes increasingly crucial as the number of EVs on the road rises, impacting the grid with their charging demands. Intelligent scheduling of charging sessions and adjustment of charging rates enable utility operators to mitigate peak demand surges, achieving a better balance between energy production and consumption. Consequently, grid stability and efficiency are enhanced. Furthermore, smart EV charging provides valuable data for energy management and consumption analysis. By collecting and analyzing data on charging patterns, energy consumption, and grid interactions, utility operators and municipalities can gain insights into EV usage trends, load profiles, and grid performance. This data serves as a basis for informed decision-making, informing infrastructure planning, policy formulation, and the development of future charging networks [35, 39, 45]. Moreover, it facilitates the integration of renewable energy sources, such as solar and wind power, by optimizing EV charging to align with the availability of clean energy. Smart EV charging offers a range of benefits for municipalities and utility operators. It allows EV owners to easily find charging stations, charge their vehicles faster and safer, and save costs through network incentives and benefits. Smart charging also helps enhance grid stability by enabling remote control and synchronization of charging with grid availability [46]. Below are summarized benefits for businesses and charging operators:

- Remote monitoring and control of EV charging.
- Access to usage statistics and data.
- Management and monitoring of charging station issues.
- Convenient ability to modify pricing packages and charging station information and seamless energy metering.
- Efficient billing processes, both on-site and off-site.
- Enhanced billing options, such as pay-as-you-use or subscriptions.
- Electricity consumption management at stations, particularly useful for managing peak and off-peak pricing.
- Improved asset function and integrity management.
- Extension of asset life span.

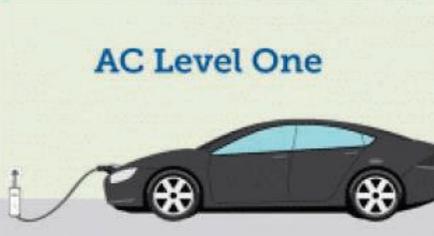

**Fig. 7.** AC Charging and DC Fast Charging

### 4.7. Research and Development (R&D)

The progression of electric vehicles (EVs) has been primarily fueled by research and development (R&D) in power electronics, electric motors, and battery storage systems. These research areas enable the creation of electric drive technologies or powertrains, that can match the performance of traditional car propulsion systems [47]. EV R&D aims to achieve several key objectives:

- Reduction in cost, weight, and volume of critical components, including energy storage systems.
- Enhancements in EV performance, efficiency, and reliability.
- Development of innovative, scalable, and modular designs for EVs.
- Improvements in manufacturing processes to streamline production.
- Acceleration of commercialization, ensuring EVs reach the market faster.

As the automotive industry continues its electrification journey, various stakeholders, including car designers, manufacturers, charging service providers, and the power industry, collaborate to standardize components and infrastructure, ensuring the safe operation and maintenance of EVs [7, 11, 39, 48]. Currently, three major areas receive considerable attention in terms of standardization:

1. *EV Batteries:*
   - Focus on battery range, weight, and size considerations.
   - Emphasis on functional and electrical safety of battery systems.
   - Environmental and performance testing to ensure reliability and compliance.

2. *EV Charging:*
   - Development of communication protocols to enable seamless interaction between charging infrastructure and EVs.
   - Addressing market-specific requirements to ensure compatibility and interoperability.
   - Advancements in wireless and inductive charging technologies.

3. *EV Electronics and Components:*
   - Adherence to ISO and IEC standards for EVs.
   - Standardization of components such as inverters, converters, and onboard chargers.
   - Specification of connectors, plugs, charging cables, and related components.

By establishing standards in these areas, the industry aims to promote consistency, interoperability, and safety across the entire EV ecosystem. This fosters the widespread adoption of EVs and facilitates a smooth charging experience for EV owners. EV R&D focuses on power electronics, electric motors, and battery storage systems to enhance EV performance. Standardization efforts target EV batteries, charging infrastructure, and electronics/components to ensure safety, interoperability, and market compatibility. Through research advancements and standards, the industry is driving the growth and acceptance of electric vehicles globally [49].

### 4.8. Growth Factors – Drivers and Resistors

The growth of EVs is propelled by technological advancements, favorable raw material prices, improvements in energy and charging infrastructure, incentives, and policies, and market readiness. However, challenges arise from market uncertainties, infrastructure limitations, higher upfront costs, limited model availability, resistance from industry stakeholders, and a lack of public awareness regarding EV benefits [47, 48].

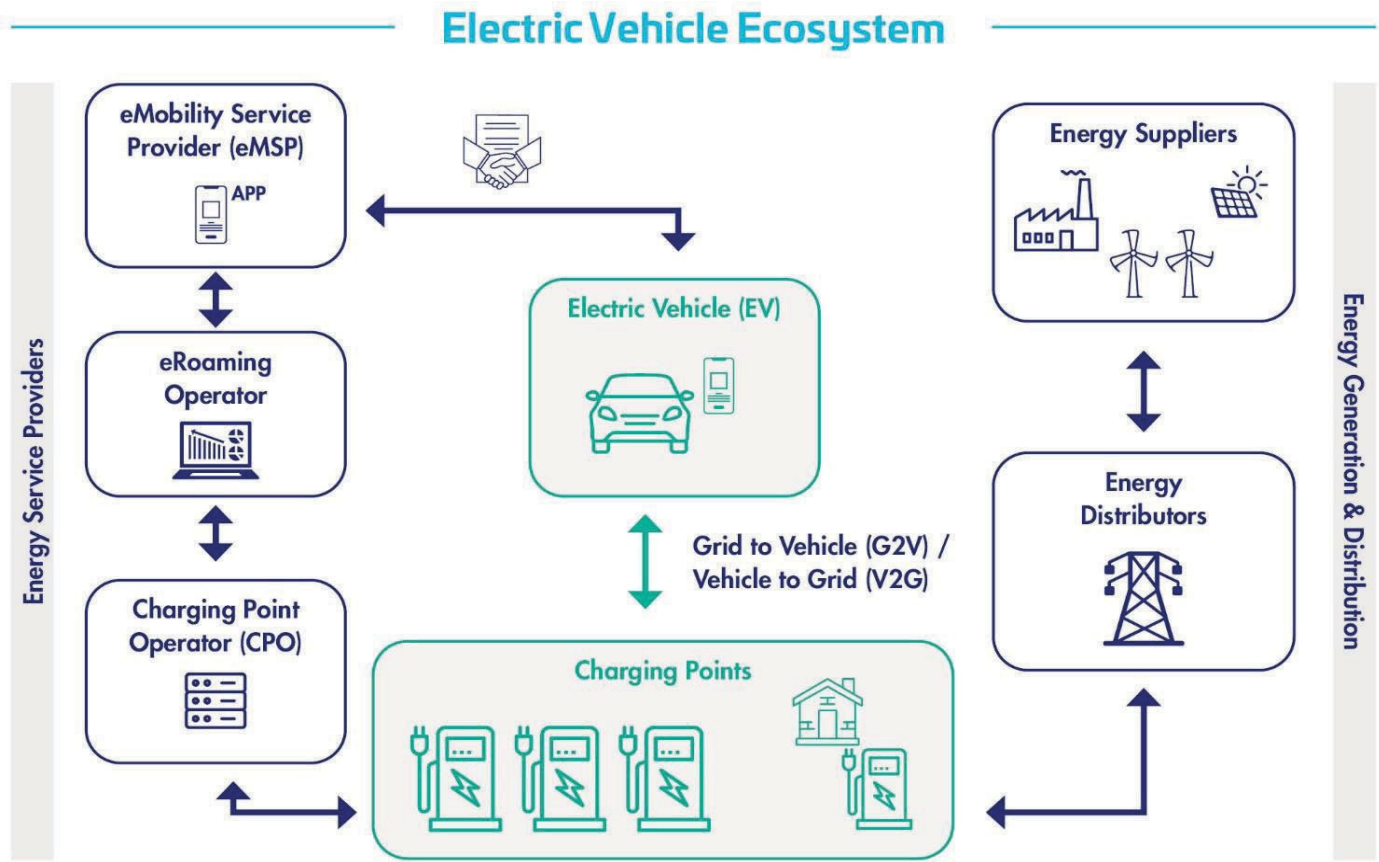

**Fig. 8.** A Typical EV Ecosystem and Smart Charging

The drivers and resistors of EV growth can be categorized as follows:
1.  Drivers:
*Technological Advancements:*
- Advancements in battery technology, resulting in reduced production costs.
- Enhanced energy density to increase range and efficiency.
- Improved chargers, reducing battery charging time and driving global adoption.

*Price of Raw Materials:*
- Reduction in the price of critical raw materials such as Cobalt, lithium, silicon, and other battery and charging components.
- Lower manufacturing costs and sales prices for EVs.

*Energy and Charging Infrastructure:*
- Progress in power stability, availability, generation, and transmission, facilitating the deployment of EV charging infrastructure on a broader scale.
- Availability of power suitable for fast charging, influencing EV adoption, especially in developing nations.

*Incentives and Policies:*
- Purchase subsidies, and trade-in incentives.
- Financial support for infrastructure development.
- Tax breaks and credits for EV purchases.
- Standards and mandates for hardware and mobility services.
- Import and export regulations.
- Emission policies and sustainable development goals/targets.

*Market Readiness:*
- Implementation of policies and incentives indicating market readiness and attracting investors' interest.
- Governments' environmental and sustainability objectives are supported by policy and political determination.
- Public perception, necessitating educational initiatives by manufacturers and stakeholders, as well as a wide range of car types, functions, and designs to promote adoption.

2.  Resistors:
*Market Conditions and Policies:*
- Uncertainty in market conditions and policy frameworks affecting the growth of EVs.

*Charging Infrastructure:*
- Insufficient stable power supply and charging infrastructure in certain regions, hindering EV adoption.

*Cost Factors:*
- Higher initial costs compared to internal combustion engine vehicles.
- Limited variety and options for EV models in certain markets.

*Industry Resistance:*
- Resistance from established stakeholders in the traditional automotive industry.

*Awareness and Understanding:*
- Limited public awareness and understanding of the benefits and capabilities of EVs.

*4.9. EV-Related Policies and Top Makers*

Governmental policies, both at international and local levels, have a significant influence on EV adoption. Major EV markets have implemented effective policies to facilitate the transition to electric mobility. Initially, fiscal incentives and subsidies played a vital role in supporting EV manufacturers and users due to high manufacturing costs. These incentives included purchase subsidies, financing options, scrappage bonuses, infrastructural development financing, and federal tax credits as shown in Figure 9. However, technological advancements in battery efficiency, battery charging, and energy infrastructure have led to a reduction in the overall cost of manufacturing, purchasing, and maintaining EVs [49, 50].

Consequently, there is a noticeable shift from direct subsidies to policy approaches that rely more on regulatory and structural measures. These measures encompass zero-emission vehicle mandates and fuel economy standards, which provide clear and long-term signals to the automotive industry and consumers, promoting the economically sustainable transition to EVs. It is essential to tailor policies to support the specific market transition in each locality. The USA, Germany, France, South Korea, Japan, and China are actively engaged in the EV sector as shown in Figure 10. While many of these manufacturers are established carmakers, there are also new companies solely dedicated to EVs. As the adoption of EVs continues to grow, it is anticipated that other regions around the world will adopt supportive policies and expedite the shift toward electric mobility [49].

| | | Canada | China | EU | India | Japan | US |
|---|---|---|---|---|---|---|---|
| Regulations (Vehicles) | ZEV Mandate | ✓* | ✓ | | | | ✓* |
| | Fuel Economy Standards | ✓ | ✓ | ✓ | ✓ | ✓ | ✓ |
| Incentives (Vehicles) | Fiscal Incentives | ✓ | ✓ | ✓ | ✓ | | ✓ |
| Targets (Vehicles) | ZEV Mandate | ✓ | ✓ | ✓ | ✓ | ✓ | ✓* |
| Industrial Policies | Subsidy | ✓ | ✓ | | | ✓ | |
| Regulations (Chargers) | Hardware Standards** | ✓ | ✓ | ✓ | ✓ | ✓ | ✓ |
| | Building Regulations | ✓* | ✓* | ✓ | ✓ | | ✓* |
| Incentives (Chargers) | Fiscal Incentives | ✓ | ✓ | ✓ | ✓ | | ✓* |
| Targets (Vehicles) | | ✓ | ✓ | ✓ | ✓ | ✓ | ✓* |

**Fig. 9.** EV-Related Policies in Selected Regions

*4.10. EV-Related Policies and Top Makers*

In Nigeria, several laws and regulatory institutions are relevant to various aspects of the country's electricity industry and product standards [51]. These include:

1. Nigerian Electricity Regulatory Commission (NERC): NERC serves as the regulator of the electricity industry in Nigeria. Its main responsibility is to enforce the Electric Power Sector Reform Act (EPSRA) and oversee related matters.
2. Standard Organization of Nigeria (SON): SON issues the Mandatory Conformity Assessment Program (MANCAP) Certificate for locally manufactured products in Nigeria. This ensures that these products conform to the relevant Nigerian Industrial Standards (NIS) before being sold in Nigeria or exported. SON also issues the Standards Organization of Nigeria Conformity Assessment Program (SONCAP) Certificate for imported products, including components and equipment used in power system installations.
3. National Office for Technology Acquisition and Promotion (NOTAP): NOTAP is responsible for registering contracts related to the transfer of foreign technology to Nigerian parties. It also handles agreements involving trademarks, patented inventions, technical expertise, engineering services, machinery, and plant supply, among others.
4. Nigerian Electricity Management Services Agency (NEMSA): NEMSA is responsible for conducting electrical inspectorate services in Nigeria's electricity supply industry. It ensures that major electrical materials and equipment used in the country meet the required quality and standards.
5. National Agency for Food and Drug Administration and Control (NAFDAC): NAFDAC is responsible for regulating and controlling the importation, exportation, manufacture, advertisement, distribution, sale, and use of various products, including chemicals. If chemicals are imported, manufactured, or used in the manufacturing process, NAFDAC permits are required.

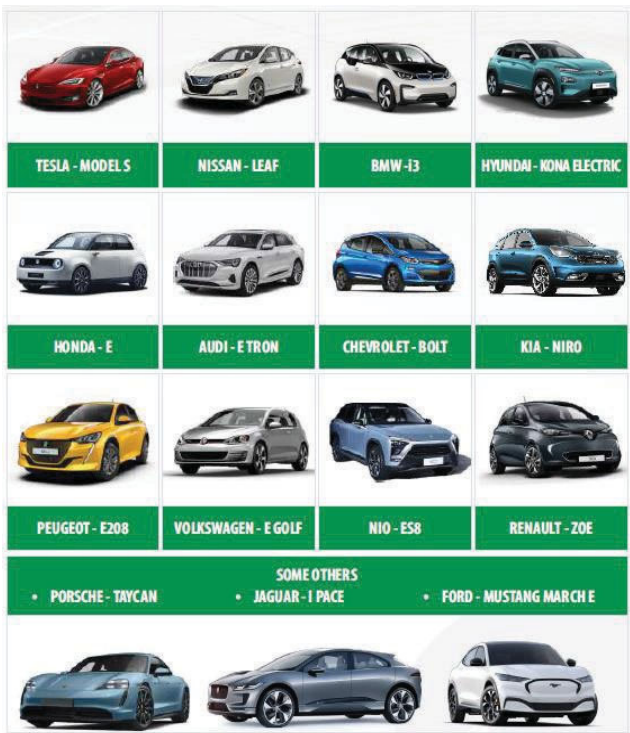

**Fig. 10.** Top EV Models and Manufacturers

These laws and regulatory institutions play crucial roles in overseeing and ensuring compliance with standards, safety, and quality in Nigeria's electricity industry and product manufacturing sector. For electric vehicles (EVs) in Nigeria, some additional authorizations and institutions may be relevant [51]. These include:

1. Environmental Impact Assessment (EIA) Certificate: The Federal Ministry of Environment issues this certificate to confirm that an adequate assessment of the environmental impact has been conducted for EV or battery manufacturing projects or charging station operations.
2. NEMSA Certificate: For the deployment of EV components, batteries, and charging stations, a NEMSA certificate is required. It is obtained from the Nigerian Electricity Management Services Agency.
3. Building & Construction Permits: Various state land and physical planning agencies issue permits for construction activities at EV project sites, including charging stations.
4. Factories License: To operate a factory for EV manufacturing, a license is needed from the Director of Factories in the Ministry of Labour.
5. NAFDAC Certificate: The importation or use of industrial chemicals for manufacturing EVs, batteries, or charging stations necessitates a certificate from the National Agency for Food and Drug Administration and Control.
6. NESREA: For the importation of new electrical/electronic equipment and proper waste management during EV project construction, compliance with regulations from the National Environmental Standards Regulation Enforcement Agency is required.
7. Import Related Permits: Importing components for EVs, batteries, and charging stations necessitates various permits, including those issued by the Central Bank of Nigeria and the Standards Organization of Nigeria.
8. Import Clearance Certificate: To import fully assembled generators, knocked-down parts for domestic assembly, or spare parts, an import clearance certificate from the Nigerian Customs Service (NCS) is required.

These authorizations and institutions ensure compliance, environmental assessment, quality standards, and proper management throughout the lifecycle of EVs in Nigeria [52].

**5. Conclusion**

In conclusion, this paper emphasizes the crucial role of all stakeholders and non-stakeholder actors in Nigeria in exploring opportunities within the renewable energy sector, particularly in relation to the adoption of electric vehicles. The evolving energy landscape, influenced by volatile global oil prices, evolving renewable energy policies, and changing interests of financial institutions, necessitates the diversification efforts of all stakeholders and non-stakeholder actors in Nigeria.

The increasing popularity of electric vehicles disrupts the demand for fossil fuels, further underscoring the urgency for diversification. Nigeria, with its abundant solar energy resources and dynamic economic development, presents a promising environment for private entities seeking to pursue environmentally sustainable initiatives. Moreover, there is a growing interest among financiers to invest in renewable energy projects across Africa, including Nigeria.

The Nigerian government's focus on solar power projects for rural and institutional electrification enhances the sector's potential for growth. By embracing environmentally sustainable initiatives and harnessing Nigeria's solar energy potential, all stakeholders and non-stakeholder actors can significantly contribute to shaping a cleaner and more sustainable future. Leveraging the support of financiers and the government will play a vital role in facilitating these efforts, ultimately leading to the realization of feasible electric vehicle adoption in Nigeria.

**Declaration of competing interest**

The authors declare that they have no known competing financial interests or personal relationships that could have appeared to influence the work reported in this paper.

**Acknowledgments**

We acknowledge and appreciate Oando Plc for the release of the abridged version report on the pre-feasibility of electric vehicles in Nigeria. This work was extracted from the report which was initially supported and funded by the Oando Plc Industry. The views expressed in this paper do not directly or necessarily reflect Oando Plc's version.